\documentclass[iop]{emulateapj}
\usepackage{wrapfig}
\usepackage{amsmath}
\usepackage{natbib}
\usepackage{color}
\usepackage [autostyle, english = american]{csquotes}
\MakeOuterQuote{"}
\newcommand{\mcrit}{M_\text{crit}}
\newcommand{\mcore}{M_\text{core}}
\newcommand{\mearth}{M_{\Earth}}
\newcommand{\mgas}{M_{\text{gas}}}
\newcommand{\water}{\text{H}_2 \text{O}}
\newcommand{\cotwo}{\text{CO}_2}
\newcommand{\methane}{\text{CH}_4}
\newcommand{\ammonia}{\text{NH}_3}
\newcommand{\vhill}{v_H}
\newcommand{\msun}{M_{\sun}}

\begin{document}

\submitted{Accepted to AJ: 21 July 2017}
\title{The Formation of Uranus and Neptune: Fine Tuning in Core Accretion}
\shorttitle{Dynamical Upheaval in Ice Giant Formation} 
\shortauthors{Frelikh \& Murray-Clay}
 
\author{Renata Frelikh \altaffilmark{1} and Ruth A. Murray-Clay \altaffilmark{1}}
\altaffiltext{1}{Department of Astronomy and Astrophysics, University of California, Santa Cruz, CA 95064; rfrelikh@ucsc.edu}


\begin{abstract}
Uranus and Neptune are ice giants with $\sim$ 15\% atmospheres by mass, placing them in an intermediate category between rocky planets and gas giants. These atmospheres are too massive to have been primarily outgassed, yet they never underwent runaway gas accretion. The ice giants never reached critical core mass ($\mcrit$) in a full gas disk, yet their cores are $\gtrsim \mcrit$, suggesting that their envelopes were mainly accreted at the end of the disk lifetime. Pebble accretion calls into question traditional slow atmospheric growth during this phase. We show that the full-sized ice giants predominantly accreted gas from a disk depleted by at least a factor of $\sim 100$. Such a disk dissipates in $ \lesssim 10^5$ years. Why would both cores stay sub-critical for the entire $\sim$ Myr disk lifetime, only to reach $M_\text{crit}$ in the final $10^5$ years? This is fine tuned. Ice giants in the outer disk have atmospheric mass fractions comparable to the disk gas-to-solid ratio during the bulk of their gas accretion. This point in disk evolution coincides with a dynamical upheaval: the gas loses its ability to efficiently damp the cores' random velocities, allowing them to be gravitationally excited by Jupiter and Saturn. We suggest that the ice giants' cores began growing on closer-in orbits (staying sub-critical), and migrated out during this dynamical instability. There, their orbits circularized after accreting much of their mass in solids. Finally, they accreted their envelopes from a depleted nebula, where the sparseness of feeding zone gas prevented runaway.

\end{abstract}


\section{Introduction}
Uranus and Neptune, ice giants with masses 14.5 $\mearth$ and 17 $\mearth$, respectively, consist of cores of rock and ice, surrounded by hydrogen and helium envelopes. Models of the planets' interiors indicate that these atmospheres comprise about 12-14\% of Uranus and 14-16\% of Neptune by mass \citep{nettleman2013}. This places the ice giants in an intermediate category between the rocky planets, with atmospheres likely dominated by outgassing, and the gas giants, whose atmospheres are a result of runaway gas accretion from the nebula.

A simple scaling argument suggests that the atmospheres of Uranus and Neptune were primarily accreted from the nebula, rather than outgassed from the cores. Suppose that the hydrogen was originally bound in the core as $\water$. Releasing 2 $\mearth$ in hydrogen would require a starting mass of 18 $\mearth$ of water, which is comparable to the entire mass of the core. This is unfeasible, since after outgassing the core would almost entirely consist of oxygen. (D. Stevenson - private communication). In principle, 2 $\mearth$ of hydrogen can be released from 11 $\mearth$ of $\text{NH}_3$ or 8 $\mearth$ of $\methane$. However, observations show that the carbon in protoplanetary disks is primarily bound in CO and $\cotwo$ \citep{oberg2008, pontoppidan2008}, and that only 10-12\% of the nitrogen is in ammonia ices \citep{bottinelli}. This can be understood by a theoretical argument: the disk abundances are inherited from the interstellar medium, where the carbon is in $\cotwo$, and CO \citep{draine2003}, and the nitrogen is in $\text{N}_2$. Though protoplanetary disk conditions favor $\methane$ and $\text{NH}_3$, the reaction timescales for the conversion of nitrogen to ammonia and carbon monoxide to methane are too long compared to the disk lifetime \citep{lodders2003}. 

For our argument, we are interested in maximizing the potential contributions of $\methane$ and $\ammonia$ to the core composition at the ice giants' disk locations, which are comparable to or beyond the snowline locations of $\methane$ ($\simeq 4$ AU) and $\ammonia$ ($\simeq 23$ AU) \citep{piso2014}. Since the presence of large amounts of pure $\text{CH}_4$ and $\text{NH}_3$ in the disk is unlikely, we can consider an $\water$ planetary core with $\methane$ and $\ammonia$ mixed in at abundances corresponding to the maximum observed abundances in a disk. We define $n_\text{X}$ as the number density of species X. Using the maximum observed abundances of $n_{\methane, \text{max}} = 0.13 n_{\water}$ \citep{oberg2008}, and $n_{\ammonia, \text{max}} = 0.15 n_{\water}$ \citep{bottinelli}, outgassing 2 $\mearth$ of $\text{H}_2$ requires a starting mass of $\sim$ 17 $\mearth$ of a mixture of $\water$, $\methane$, and $\ammonia$. Forming a $\sim 2$ $\mearth$ atmosphere would require releasing all of the hydrogen from the core, leaving behind a $\sim 15$ $\mearth$ C/N/O core. Such an outcome is implausible: the ice giants' cores almost certainly contain hydrogen. A 100\% dissociation efficiency is highly unlikely; moreover, most of the hydrogen would remain mixed in the core, as it would not phase separate \citep[e.g.][]{2015ApJ...806..228S}. We conclude that the hydrogen envelopes of Uranus and Neptune cannot be outgassed.

The standard scenario for the formation of planets with a significant gas component is core growth by planetesimal accretion, followed by gas accretion \citep{pollack} once a critical core mass is reached. For the purposes of this paper, we define the critical core mass ($\mcrit$) to be the mass required for the protoplanetary core to be able to accrete a gas envelope with a mass $\mgas$ comparable to the core mass by a given time (generally, the gas disk lifetime). Once $\mcore=\mgas$, the atmosphere starts to contribute substantially to the gravitational potential of the planet, triggering runaway gas accretion by nucleated instability \citep[e.g.][]{pollack, rafikov2011}. We note that $\mcrit$ is defined in two ways in the literature: the steady state solution described by \citet{rafikov2006}, and the time evolving solution \citep{pollack}. The \citet{rafikov2006} solution considers a core with an accretion luminosity that is high enough that its gas envelope is able to reach equilibrium on a short timescale compared to the disk lifetime. In this case, the envelope achieves a pseudo-steady state, with radiative losses balancing heating by accreting planetesimals. So, $\mcrit$ is the core mass that is able to hold onto its own mass in gas, while continually being injected with energy from accretion. The planet undergoes runaway gas accretion if the core is able to grow to $\mcrit$ in the lifetime of the disk. On the other hand, the \citet{pollack} solution requires accretion to halt (or slow substantially) once the supply of planetesimals is depleted, allowing the envelope to accumulate gas as it gradually cools and undergoes Kelvin-Helmholtz (KH) contraction. If the core is able to accumulate its own mass in gas within the lifetime of the full disk, it undergoes runaway gas accretion and becomes a gas giant, with $\mcrit$ being the core mass at the point when runaway is triggered. When not otherwise specified, below we use $\mcrit$ generically to mean the mass at which a core is able to accrete $\mgas=\mcore$ by any mechanism.

What separates the ice giants from the gas giants is their formation history: during core growth, the former never exceeded $\mcrit$ in a full gas disk. While sub-critical, the relatively small cores were unable to accrete much gas, with the core mass as a conservative upper limit to the mass of the atmosphere at any given time before the planets reached critical core mass. On the other hand, if given sufficient time to accumulate gas, a super-critical core is able to accrete most of the gas in its feeding zone, which in a full disk is typically sufficient to produce a gas giant.


The cores of Uranus and Neptune exceed reasonable estimates of the critical core mass. One can imagine several scenarios to prevent runaway. A high enough planetesimal accretion rate can cause the critical core mass to exceed the current core masses of the ice giants (Section 2). However, the high accretion rate must be sustained for the bulk of the gas disk lifetime, which is unlikely. As an alternative, the standard explanation of \citet{pollack} avoids fine tuning by proposing the existence of a stage of slow gas and planetesimal accretion that persists for a couple Myr. In this stage, the composition of the ice giants remains similar to their present-day makeup for a significant fraction of the gas disk lifetime. However, pebble accretion \citep{lambrechts2012, ormel2010} makes this explanation newly problematic, as we will show in section 2.2. Finally, our ice giants could have reached their final core masses in a largely depleted gas disk, when there was just enough gas left in their feeding zones to accrete their intermediate-sized atmospheres. We show in Section 3 that this corresponds to a disk mass of at most $\lesssim 0.01$ times its original mass, and show in Section 4 that this depleted disk gas dissipates fairly quickly ($\lesssim 10^5$ years).

Once the gas is depleted to $\simeq 0.01$ of the full gas disk, the mass of gas in the disk becomes comparable to the mass of solids. A solid-to-gas ratio of order unity is a known factor for dynamical instability. This dynamical instability is triggered when the gas becomes so depleted that it can no longer effectively damp the protoplanetary cores' random velocities. In Section 5, we discuss a possible solution for the fine tuning problem: there is a dynamical reason for the ice giants to accrete most of their mass in gas at the end of the gas disk lifetime.

\section{Avoiding runaway}
We now discuss the standard ways to avoid runaway. In 2.1, we discuss the possibility that the growing protoplanetary core could have accreted planetesimals so quickly that the entropy deposited into the envelope was enough to prevent collapse. In other words, $\mcrit$ was larger than the current mass of the ice giants throughout the gas disk lifetime. In this fast accretion regime, the atmosphere satisfies a static solution, where the core and the envelope have reached a state of thermal equilibrium, with the energy in balancing the energy out \citep{rafikov2006}. In 2.2, we consider the classical argument of \citet{pollack}, in which the ice giants' core growth effectively halts at a mass lower than would be necessary to quickly accrete an envelope. In this dynamic solution, which differs from the scenario discussed in 2.1 in the sense that the atmosphere never reaches a steady state, the atmosphere instead progresses through a period of slow gas accretion limited by the cooling rate.



\subsection{Can the cores stay sub-critical due to fast planetesimal accretion?}

\begin{figure}[h!]
 \centering
   \includegraphics[width=0.47\textwidth]{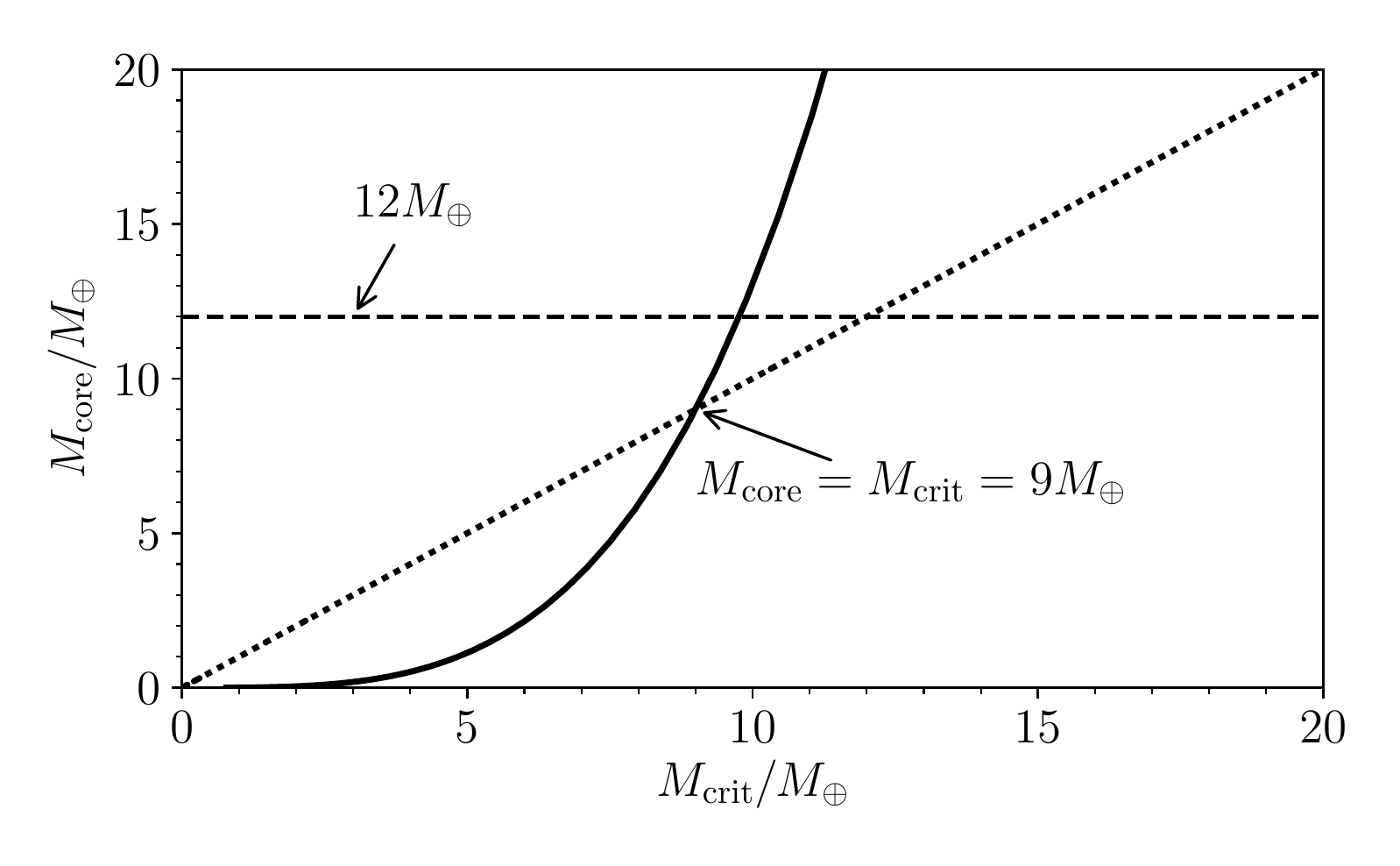}
   \caption{Is it possible to keep the core sub-critical by accreting at a constant $\dot{M}$ for the $\sim$ 3Myr gas disk lifetime? We define $M_\text{core} = \dot{M}\tau_{\text{disk}}$ as the size the core can attain in the lifetime of the disk, assuming it is accreting at the rate $\dot{M}$ associated with $\mcrit$ (Equation \ref{m_crit}) for the entire disk lifetime (solid line). The cutoff mass for accretion to keep the core sub-critical is $M_{\text{core}} = \mcrit$ (dotted line). At a higher accretion rate, the cores cannot stay sub-critical because the accretion rate required to prevent runaway would cause the core to grow larger than its assumed size.
   \newline
   The values of $M_{\text{core}}(\mcrit)$ are calculated for $\rho_c = 3\text{g cm}^{-3}$ (the typical core mean density for Uranus and Neptune from \citet{nettleman2013}), $\kappa_0 = 0.1 \text{cm}^2\text{g}^{-1}$, and $\mu = 2.3 m_p$ (mean molecular mass of $\text{H}_2/\text{He}$ gas).}
 \end{figure}

As planetesimals are accreted onto a growing core, their gravitational potential energy is released near the surface with an accretion luminosity $L = GM_p \dot{M} /R$. This increases the temperature at the base of the atmosphere and puffs it up, thus decreasing the total atmospheric mass compared to that of an otherwise identical non-accreting core, and increasing the amount of energy that must be radiated away to bind additional gas. 

In Equation \ref{m_crit}, we express the functional dependence of the critical core mass on the planetesimal accretion rate ($\dot{M}$) and the gas opacity in the outer radiative layer of the atmosphere ($\kappa_0$) from \citet{rafikov2011}. We note that this is the applicable value for $M_{\text{crit}}$ in the high accretion rate limit, for which energy in balances energy out.
\begin{equation} \label{m_crit}
M_\text{crit} = \left( \left(\frac{k_B}{\mu} \right)^4 \left(\frac{4\pi \rho_c}{3}\right)^{1/3}\frac{\kappa_0 \dot{M}}{\zeta_0 \sigma G^3} \right)^s,
\end{equation}
with constants of order unity $s = \frac{3}{7+3\delta}$ and $\delta = 1.2$, a constant $\zeta_0 = 4\text{x}10^{-32} \text{ g}^{-\delta}$ that depends on $\delta$, the mean molecular mass $\mu$, the Boltzmann constant $k_B = 1.38\text{x}10^{-16} \text{cm}^2 \text{g } \text{s}^{-2} \text{K}^{-1}$, the gravitational constant $G = 6.67\text{x}10^{-8}\text{cm}^3 \text{g}^{-1} \text{s}^{-2}$, the Stefan-Boltzmann constant $\sigma = 5.67\text{x}10^{-5}\text{g } \text{s}^{-3} \text{K}^{-4}$, and the mean density of the core $\rho_c$. In this context, the critical core mass is calculated assuming steady state can be achieved. We note that a high accretion luminosity (e.g. comparable to that of a core in the fast or intermediate accretion regime of \citet{rafikov2006}) increases the total luminosity of the core, causing it to evolve toward this steady state on timescales typically shorter than the disk lifetime ($\tau \sim E/L$, where $E$ is the total energy of the envelope).

How massive can the core grow in the lifetime of the gas disk (using $\tau_\text{disk} \simeq 3$ Myr as a conservative estimate) and stay sub-critical? The maximum accretion rate that the core can sustain for the full disk lifetime without growing beyond its final size ${M}_\text{core}$ is: $\dot{M} = M_\text{core}/\tau_\text{disk}$. 

In Figure 1, we plot Equation \ref{m_crit} for $\kappa = 0.1 \text{cm}^2 \text{g}^{-1}$, a commonly assumed opacity in protoplanetary disks \citep[e.g][]{beckwith, pollack1985}. If ${M}_{\text{core}} = \dot{M}\tau_{\text{disk}} < \mcrit$, where $\dot{M}$ is the accretion rate associated with $\mcrit$, then accretion can in principle keep the core sub-critical. If accretion is not sustained at this value throughout the final stages of core growth, which in standard models is comparable to the full disk lifetime, the effective value of ${M}_{\text{crit}}$ is reduced. We may therefore calculate the maximum core mass consistent with runaway prevented by sustained accretion at the rate $\dot{M}$. We conclude from  Equation \ref{m_crit} that $\sim 10 M_{\Earth}$ is the greatest mass the core can grow to in $\tau_\text{disk}$ and remain just sub-critical. In Figure 1, the accretion rate that gives a critical core mass of $10 M_{\Earth}$ allows the core to grow to $10 M_{\Earth}$ in the lifetime of the disk. Increasing the accretion rate to above this value is not enough to prevent runaway; to the contrary, it gives rise to a super-critical core. We conclude that for our assumed values of opacity, a high accretion rate could not have kept the $\gtrsim 12 M_{\Earth}$ cores of Uranus and Neptune from reaching critical core mass.

The dependence of the critical core mass on opacity introduces a degree of uncertainty. For example, increasing the grain opacity by an order of magnitude ($\simeq$1 $\text{cm}^2\text{g}^{-1}$) doubles the critical core mass, allowing the cores to stay sub-critical for the lifetime of the disk for a limited range of accretion rates. However, this still requires fine tuning: the cores would have had to accrete at a specific rate for the entire disk lifetime.

In the discussion above, we have considered a scenario of core accretion in which a steady-state solution for the envelope exists, given a high planetesimal accretion rate $\dot{M}$. In this scenario, planetesimal accretion acts as a continuous heating source for the envelope that balances its outgoing luminosity, which prevents the envelope from contracting and sustains it in a steady state. It is plausible that the protoplanets were in this high-accretion regime at some point in their evolution, but we have shown that they could not have sustained it for their entire formation time, hence this is not a way to prevent collapse. In 2.2, we discuss an alternate way the ice giants could have avoided runaway by considering growth with a low planetesimal accretion rate, which can be sustained for a significant fraction of the gas disk lifetime. If $\dot{M}$ is slow enough, in principle the cores can be limited to small enough masses that they wouldn't be able to quickly accrete the amounts of gas necessary to trigger runaway gas accretion.

\subsection{Discussion of the Classical Core Accretion Scenario}
We now discuss the classical work of \citet{pollack}, which models the formation of the giant planets with a 3-phase process that includes the concurrent accretion of gas and solids. We focus on the key assumptions of this model, and refer to the original work for a detailed description of the model. 

During Phase 1, the core grows by planetesimal accretion with a rate $\dot{M}_{\text{solids}}$ until it depletes the bulk of the material in its feeding zone with no replenishment from the outside, reaching isolation mass at a time $\tau_{\text{iso}}$. This model assumes that the protoplanetary core carves a gap in the planetesimal disk, causing core growth to halt at roughly the local isolation mass. The core starts accreting an envelope once the core is massive enough that its escape speed is greater than the local sound speed of the gas. The envelope is required to cool and contract for it to grow, and its growth rate can be stalled by external heating. For example, the envelope cannot cool as efficiently via KH contraction if planetesimal accretion deposits extra energy into the envelope, compared to the case with no planetesimal heating: the planetesimal accretion luminosity will be part (if not the main component) of the planet's total outgoing luminosity. This model relies on the planetesimal accretion luminosity to be initially large, and then to drop to less than the KH luminosity of the envelope. This can be achieved by cutting off the supply of planetesimals by growing the core to isolation mass, assuming no radial drift of solids through the gas. The key point is that accretion cuts off while the core remains low enough mass to take extra time to accrete gas.

However, this is problematic: observations have shown the existence of pebbles in the outer disk \citep{Andrews}, which are mm/cm-sized particles that are well-coupled to the gas. These particles may drift into the feeding zone with the gas and rapidly accrete onto the core via pebble accretion \citep{ormel2010,lambrechts2012,rosenthal2017}, which has been shown to be extremely efficient in forming cores even in the outer disk. 

Rapid accretion of mm-cm sized planetesimals (known as pebbles) by gas-assisted growth can reduce the growth timescales sufficiently to allow giant planets to form at wide orbital distances. We define the Hill radius to be the distance from the planet at which the gravity of the planet dominates over the tidal gravity of the Sun:

\begin{equation} \label{HR}
r_{H} = a\left(\frac{M_{c}}{3M_{\sun}}\right)^{1/3},
\end{equation}
where $M_{c}$ is the mass of the core, $M_{\sun}$ is the mass of the Sun, and $a$ is the semimajor axis of the planetary orbit. Under the most favorable conditions to gas-assisted growth, pebbles are able to accrete once they enter the Hill sphere, increasing the cross-section for accretion significantly from the classical one derived for gravitational focusing. For a review of classical growth rates, see \citet{gls}.

The most favorable velocity at which the accreting particles approach the core is given by the Hill velocity, $\vhill = \Omega r_H$ ("shear-dominated regime" \citet{gls}, section 2.2). This is an appropriate value for low dispersion velocities, i.e. $\lesssim \vhill$, for which accretion is 2-dimensional and particles approach the core due to Keplerian shear. In the absence of damping, the cores themselves excite nearby particles to $\sim \vhill$. We note for reference that gas can both damp the dispersion velocities of particles and - when turbulent - kick them to higher velocities. Which effect dominates depends on particle size. According to pebble accretion, in the most favorable case, the collision rate is equal to the Hill entry rate. Therefore, the cross-section for accretion from a thin disk of particles is $\sigma_{\text{acc}} \sim H r_H$, where $H$ is the particle scale height. For large ($\gtrsim M_{\Earth}$) cores accreting pebbles under typical disk conditions, the maximum growth rate (see \citet{rosenthal2017}), assuming the existence of pebbles coupled to the gas in the outer disk, is then:
\begin{equation} \label{pebble}
    \dot{M} = \rho \sigma_{\text{acc}} v_\text{rel} = \frac{\Sigma}{H} r_H H r_H \Omega = \Sigma {r_H}^2 \Omega,
\end{equation}
where $v_\text{rel}$ is the relative velocity between the pebbles and the core, $\rho$ and $\Sigma$ are the respective volume and surface mass densities of solids in the disk, and $\Omega$ is the local Keplerian frequency. The cores cannot stay at 12 $M_\earth$ for very long: the timescale for such a core to double its mass via pebble accretion can be as short as $10^5$ years (Equation \ref{pebble}) at the current distance of Neptune in a disk with a Minimum Mass Solar Nebula (MMSN) solid surface density profile \citep{1981PThPS..70...35H}. We note that the minimum timescale for gas-assisted growth does not depend on the surface density of gas, which primarily affects the sizes of the pebbles accreted \citep{rosenthal2017}, and that increasing the surface density in solids can significantly decrease this timescale. The fast growth rates predicted by pebble accretion for large cores and the fact that radial drift through the disk replenishes the planetesimal supply are problematic for the original core accretion interpretation for ice giants, which relies on core growth halting at the isolation mass. In other words, having a roughly fixed-mass core over timescales comparable to the disk lifetime is needed for slow gas accumulation to explain the properties of the ice giants' envelopes without fine tuning, and gas-assisted growth suggests that $\sim 12 \mearth$ cores should grow rapidly.

Pebble isolation \citep{lambrechts2014} can possibly save the original interpretation: pebble accretion can be halted once the core reaches a mass at which the core can gravitationally perturb the gas disk such that entrainment in gas flowing around the core stops the drift of pebbles onto the core. We now estimate the pebble isolation mass (see \citet{rosenthal2017} for a detailed discussion). We define the Bondi radius as the radius at which the gravitational potential energy of the core equals the thermal energy of the gas: $R_{B} = 2GM_{c}/{c_{s}^2}$, where $c_{s}$ is the sound speed of the gas and $M_{c}$ is the core mass. The pebble isolation mass is the core mass required for the Bondi radius to surpass the Hill radius:
\begin{equation} \label{pebiso}
M_{\text{iso,peb}} \simeq \left(\frac{a^3}{24M_{\sun}G^3}\right)^{1/2}c_{\text{s}}^3 \simeq 27 M_{\earth} \left(\frac{a}{20 \text{AU}}\right)^{3/4},
\end{equation}
using a temperature profile of $T = 300 (a/\text{AU})^{-1/2}$ K appropriate for the MMSN in estimating the sound speed. Our estimate of $M_{\text{iso,peb}}$ at ice-giant distances is close enough (to within an order of magnitude) to the current core masses of Uranus and Neptune that halting accretion of pebble-sized objects via this mechanism cannot be ruled out with certainty. Alternatively, pebble accretion can halt if the supply of pebbles in the outer disk is cut off, which, based on the observations of \citet{Andrews}, we suppose is a less likely explanation.

Assuming core growth can be halted at a mass low enough that the planet cannot quickly accumulate a mass in gas large enough to trigger runaway gas accretion, the planet enters a long-lived phase of simultaneous, slow accretion of solids and gas (Phase 2), which ends when the core accumulates a mass in gas comparable to its own mass. We note that the length of Phase 2 depends strongly on the opacity, which is uncertain. Work including realistic opacities by \citet{movshovitz} and \citet{mordasini} suggests that the values for the opacities used in the classical \citet{pollack} simulations were too high, suggesting that it is even worth considering grain-free models in giant planet formation. Additional models including protoplanetary disk opacities by \citet{piso2015} showed that taking into account grain growth lowered the envelope opacity, yielding significantly faster runaway accretion timescales. 

Finally, for this 3-phase scenario to work, the timescale to accrete the $\sim 2 M_{\Earth}$ atmospheres of Uranus and Neptune would have to be comparable to or greater than $\tau_{\text{disk}}-\tau_{\text{iso}}$, which is a couple Myr. Furthermore, Jupiter and Saturn must have reached runaway in the $\sim$ 3 Myr lifetime of the gas disk, which does not happen for the high opacities used in \citet{pollack}. We note that the timescale for KH contraction (and hence the critical core mass for low planetesimal accretion rates) varies as a function of distance from the star primariy due to changes in opacity. Opacities in the interstellar medium scale with temperature as $\kappa \propto T^2$, and, as shown in \citet{piso2015}, the critical core mass declines with distance from the star. More realistically, once grains have grown, the opacity typically loses its temperature dependence, and the resulting critical core mass is a weakly declining function of stellocentric distance. So, for this scenario to be correct, Jupiter and Saturn must have had only marginally enough time to achieve runaway.

Given the difficulty in constraining the timescales of envelope growth due to the uncertainties in opacities, the specific opacities required to grow the atmospheres of Jupiter and Saturn within reasonable estimates for gas disk lifetimes without turning Uranus and Neptune into gas giants, and the challenges in halting the growth of a core experiencing pebble accretion, we find it valuable to consider our proposed dynamical alternative.

\section{How Depleted is the Gas at the Onset of Gas Accretion?}

We have suggested that the fully grown cores of Uranus and Neptune could not have accreted their atmospheres from a full gas disk. Instead, the cores accreted the bulk of their gas from a depleted disk late in its lifetime. Here we calculate the maximum gas mass available for accretion in the depleted disk. We note that the atmospheric growth timescale is dominated by the last doubling timescale for the atmosphere \citep[e.g.][]{piso2014}, and that the fully grown cores were massive enough to accrete all of the gas in their respective feeding zones on short timescales, unless limited by the solid accretion luminosity (see Section 5).

To calculate how depleted the disk has to be at the onset of gas accretion to produce the observed envelope masses of the ice giants, we set the mass of gas in each feeding zone equal to the mass of each planet's atmosphere, respectively. We approximate a planetary feeding zone with an annulus at a distance $a$ (the planet's semi-major axis) from the Sun, with a width of $\Delta a$ of $\simeq$ 5 Hill radii (see \citet{hilke2014}). The Hill radius, $r_H$, is the distance from the planet at which the gravity of the planet dominates the gravity of the Sun (i.e. the distance scale relevant to the planet's ability to accrete material):
\begin{equation} \label{Hill_radius}
r_{H} = a\left(\frac{M_{c}}{3M_{\sun}}\right)^{1/3},
\end{equation}
where $M_{c}$ is the mass of the core, $M_{\sun}$ is the mass of the Sun. The atmosphere of an ice giant is not massive enough to contribute substantially to the planet's gravity, so we use the core mass instead of the total planet mass in the Hill radius definition. Thus, though not entirely self-consistent, the following is reasonable estimate of the gas feeding zone mass, given that the core dominates the gravitational potential of the planet:
\begin{equation} \label{gas_mass_a}
M_{g} \simeq 2\pi a \Delta a \Sigma_{g}(a), \text{ where}
\end{equation}
$\Sigma_{g} (a)$ is the mass surface density of gas in the depleted disk at a distance $a$ from the Sun. We assume that $\Sigma_{g} (a)$ stays roughly constant across the width of the feeding zone.
Taking $\Delta a \simeq 5 r_H$, this gives us:
\begin{equation} \label{gas_mass_b}
M_{g} \simeq 10\pi a^2\left(\frac{M_{c}}{3M_{\sun}}\right)^{1/3} \Sigma_{g} (a).
\end{equation}

We denote the surface density of the full gas disk as $\Sigma_{g, 0}$. We define a disk depletion factor: $f\equiv \Sigma_{g}/\Sigma_{g,0}$. Setting $M_{g} = M_\text{atm}$ in Equation \ref{gas_mass_b} gives us the depletion factor at the onset of gas accretion:

\begin{equation} \label{depletion_factor}
f = \frac{\Sigma_{g}}{\Sigma_{g,0}} \simeq \frac{M_\text{atm}}{M_{g,0}} \simeq \frac{M_\text{atm}}{10\pi a^2\Sigma_{g,0}}\left(\frac{3M_{\sun}}{M_c}\right)^{1/3}.
\end{equation}
This depletion factor can also be understood as the fraction of the original, full gas disk feeding zone mass, $M_{g,0}$, that ended up in the atmosphere of the planet.

We use the MMSN model's disk surface density profile \citep{1981PThPS..70...35H},
\begin{equation} \label{MMSN}
\Sigma_{g,0} = 1700 \left(\frac{a}{\text{AU}}\right)^{-3/2} \text{g cm}^{-2},
\end{equation} the current values of $a$, and total planetary gas masses from \citet{nettleman2013} to calculate that $f \simeq 0.01$ when Uranus and Neptune entered the stage of gas accretion during which their atmospheres acquired most of their mass. This calculated $f$ is an upper limit because we assumed the planets accreted a minimum amount of gas, and have neglected viscous feeding zone replenishment and any significant early-on gas accretion. 

We rewrite this criterion in a simple form. We define the solid-to-gas ratio in the full gas disk: $f_s \equiv \Sigma_{s}/\Sigma_{g,0}$. For the planet's atmospheric mass to be comparable to its gas feeding zone mass, we again set $M_{g} = M_\text{atm}$ in Equation \ref{gas_mass_b}:
\begin{equation} \label{atmo_mass}
M_\text{atm} \simeq 10\pi a^2\left(\frac{M_{c}}{3M_{\sun}}\right)^{1/3} \Sigma_{g}.
\end{equation}
So,
\begin{equation} \label{atm_to_core}
\frac{M_\text{atm}}{M_c} \simeq \frac{10\pi a^2 \Sigma_{g}}{(3M_{\Sun})^{1/3} M_{c}^{2/3}} = \frac{10\pi a^2}{(3M_{\Sun})^{1/3}}M_c^{-2/3}\Sigma_{g,0}f.
\end{equation}

The cores of Uranus and Neptune are comparable to their local isolation masses \citep[e.g.][]{goldreich2004}. Though we favor pebble accretion as the dominant mode of core growth at ice giant distances, we nonetheless find it useful to express planetary masses in units of isolation mass because it allows us to write the scaling relations in a simple form.


Classically, the isolation mass is the mass of a core that has consumed all of the solid material in its feeding zone:
\begin{equation} \label{iso_mass_a}
M_\text{iso} \simeq 2\pi a \Delta a \Sigma_s(a),
\end{equation}
where $\Sigma_s(a)$ is the disk surface mass density in solids at a distance $a$ from the Sun. Taking the width of the feeding zone in Equation \ref{iso_mass_a} to be $\Delta a \simeq 5 r_H=5a(M_\text{iso}/(3M_{\sun}))^{1/3}$, we solve for $M_\text{iso}$:
\begin{equation} \label{iso_mass_b}
M_\text{iso} \simeq \frac{(10\pi a^2 \Sigma_s)^{3/2}}{(3M_{\sun})^{1/2}}
\end{equation}

Then,
\begin{equation} \label{iso_mass_c}
\frac{10\pi a^2}{(3M_{\Sun})^{1/3}} \simeq \frac{M_\text{iso}^{2/3}}{\Sigma_{s}}.
\end{equation}

Thus, from Equations \ref{atm_to_core} and \ref{iso_mass_c}, we have:

\begin{equation} \label{result}
\frac{M_\text{atm}}{M_c} \simeq f \frac{\Sigma_{g,0}}{\Sigma_s} \left(\frac{M_\text{iso}}{M_c} \right)^{2/3}  = \frac{f}{f_s} \left(\frac{M_{c}}{M_\text{iso}}\right)^{-2/3},
\end{equation}
and for $M_c \simeq M_\text{iso}$, $M_\text{atm}/M_c \simeq f/f_s$. When forming an isolation-mass scale ice giant, for the atmospheric mass to be comparable to the total planet mass, we need the surface density of gas to be comparable to the surface density of solids at the onset of gas accretion (see Figure 2).

\begin{figure}[h!]
 \centering
   \includegraphics[width=0.47\textwidth]{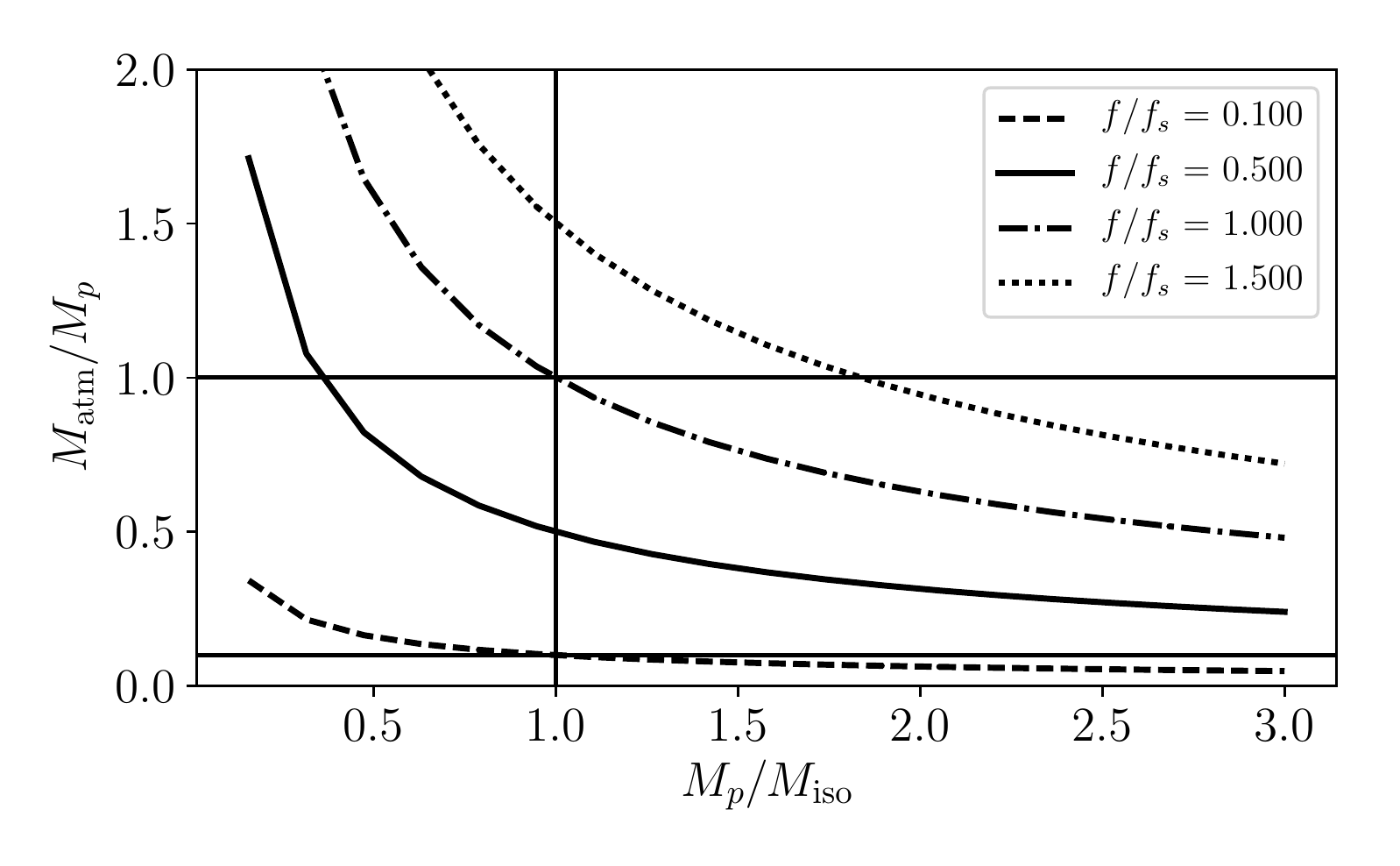}
   \caption{$M_p/M_\text{iso}$ gives the planet mass in terms of the isolation mass. $M_\text{atm}/M_p$ is the atmospheric mass fraction. Here we vary the gas-to-solid ratio, illustrating that for $\sim$ isolation-mass cores the gas-to-solid ratio at the onset of gas accretion is comparable to the final atmospheric mass fraction. The horizontal lines mark ratios of 0.1 and 1, which are reasonable for the onset of dynamical instability (see text).}
 \end{figure}
 
This point coincides with the onset of a dynamical instability in the disk: the gas can no longer damp the eccentricities and inclinations of the growing cores once the surface densities of solids and gas are comparable.

The required disk gas-to-solid ratio for the onset of instability may be less than 1. For instance, \citet{dawson2016} found in simulations of planetary embryos embedded in a gas disk that gas stopped being able to efficiently damp the embryos' eccentricities at $f = 10^{-3}$, i.e. $f/f_s \sim 0.1$. In another simulation by \citet{ford2007} of planetary cores in a planetesimal disk, a dynamical instability was triggered once the density of planetesimals dropped to roughly 0.1 times the density of big bodies. As discussed in section 5.2 of \citet{goldreich2004}, dynamical friction due to gas can be treated in the same way as planetesimal dynamical friction, replacing the small bodies' surface density with that of the gas, and the random velocity of the small bodies with the local sound speed. If the instability starts at $f/f_s \sim 0.1$, the eventual atmospheric masses will be approximately 0.1 $M_p$, which matches the ice giants' atmospheric masses quite well.

\section{How Long Can the Gas Disk Stay in a Depleted State?}
Observations of protoplanetary disks suggest that gas disks dissipate on a $\lesssim$ 3-5 Myr timescale, with most young stars stars losing their gas disks by 10 Myr \citep{haisch, alexander2014, espaillat}. Disk dispersal mechanisms include: viscous accretion onto the host star, viscous spreading, and photoevaporation driven by irradiation from the host star, as well as the local interstellar radiation field. Transitional disks, if representative of disks in the process of clearing, have lifetimes of $\sim 10\%$ of the full disk lifetime (e.g. \citet{alexander2014, espaillat}). We note that the $\sim 10\%$ transition disk fraction could instead reflect the $\sim 10\%$ frequency of planets massive enough to carve holes in the gas disk.

The predominant mode of disk dispersal can vary throughout the extent and lifetime of the disk, with the inner regions clearing due to accretion onto the star, and photoevaporation driving mass loss from the outer ones. We are interested in the outer disk, where photoevaporation can be important. The protoplanetary disk undergoes photoevaporation as it is heated by far-ultraviolet (FUV: 6 eV to 13.6 eV), extreme-ultraviolet (EUV: $>$ 13.6 eV), and X-ray photons \citep{adams2004}. FUV photons strike the disk and heat the gas to temperatures of $\approx 10^3$ K, EUV photons ionize the gas and heat it to $\approx 10^4$ K, and X-ray flux heats the gas surface layer to $\approx 10^3-10^4$ K. The critical radius is the relevant length scale to determine how far in the disk photoevaporative mass loss starts to be efficient (see Equation 1, \citet{adams2004}:
\begin{equation} \label{critical_radius}
    r_g = \frac{GM_{\Sun}\mu_\text{gas}m_p}{kT},
\end{equation}
where $\mu_\text{gas}$ is the mean molecular weight of the gas, and $m_p = 1.67\text{x}10^{-24}$ g is the proton mass. Beyond $r_g$, the local sound speed of the externally heated gas is greater than the local escape velocity from the star, allowing heated gas to escape. For a solar-mass star and atomic hydrogen gas, the critical radius for FUV is $r_g \sim 100 \text{AU}$, for EUV is $r_g \sim 10 \text{AU}$, and for X-ray the range of critical radii is $r_g \sim 10 - 100 \text{AU}$ \citep{owen2012}.

We estimate the timescale for disk dispersal due to solar EUV-driven mass loss, where the stellar irradiation drives thermal escape of gas with outflow velocities of $v_\text{flow} \simeq 10 \text{km } \text{s}^{-1}$. We check that the hydrogen gas is mostly ionized at the $\tau = 1$ layer of the disk, where $\tau$ is the optical depth to the EUV radiation. Since $\tau = N_\text{H} \sigma_\text{ion}$, the neutral column depth to $\tau = 1$ is $N_\text{H}=1/\sigma_\text{ion}$, where $\sigma_\text{ion} = 6.3 \text{x} 10^{-18}$ $\text{cm}^2$ is the photoionization cross-section for hydrogen by Lyman limit photons. The gas outflows at the critical radius $r_\text{g}$, where the heating puffs the disk up to a scale height comparable to the stellocentric distance $a$: $H \sim c_{s}/\Omega \sim v_{\text{esc}}/\Omega \sim a$. Thus, the volumetric number density of neutral hydrogen at $r_{g}$ is $n_\text{H} \sim N_\text{H}/H\sim10^3$ $\text{cm}^{-3}$. By balancing photoionization with recombination, we obtain the number density of ionized hydrogen at the $\tau = 1$ surface: $n_\text{H} \sigma \Phi_{i}/(4 \pi a^2) = \alpha{n_p}^2$, where the coefficient for radiative recombination for hydrogen at $10^4$ K is $\alpha = 2.6 \text{x} 10^{-13} \text{cm}^3 \text{s}^{-1}$, $n_p$ is the number density of ionized hydrogen/electrons (assuming charge balance), $a$ is the stellocentric distance, and $\Phi_{i}$ is the ionizing stellar flux. For a solar-mass star,
\begin{equation} \label{ionized_density}
    n_p \simeq \left(\frac{\Phi_{i}}{4\pi a^3\alpha}\right)^{1/2} \simeq 10^5 \left(\frac{\Phi_{i}}{10^{41} \text{s}^{-1}}\right)^{1/2} \text{cm}^{-3},
\end{equation}
so the assumption of ionization is correct. The the recombination timescale, $\tau_\text{rec} \simeq 1/(\alpha n_p^2) \simeq 4\text{x}10^{2} \text{s}$, is much less than the outflow timescale $\tau_\text{flow} \simeq a/v_\text{flow} \simeq 10^{8} \text{s}$, justifying the assumption of ionization equilibrium. The mass-loss rate is therefore: $\dot{M} \simeq 4 \pi a^2 m_\text{p} n_{p} v_{\text{flow}} \simeq (4 \pi a \Phi_{i} /\alpha)^{1/2} m_\text{p} v_\text{flow} \simeq 7 \text{x}10^{-10} M_{\Sun}\text{yr}^{-1}$, assuming an ionizing luminosity of $10^{41}$ photons $\text{s}^{-1}$ appropriate for T Tauri stars \citep{alexander2005}. Using the MMSN initial gas disk mass of $M = 0.01 M_{\sun}$ \citep{desch}, the gas disk already depleted by a factor of 100 disperses on a timescale of: $\tau_\text{disp} = M/\dot{M} = 10^{-4}M_{\Sun}/(7\text{ x }10^{-10}M_{\Sun}\text{yr}^{-1}) \approx 10^5 \text{ yr} $. We note that this calculation is an order-of-magnitude estimate. However, it agrees with transitional disk lifetime estimates both from statistical arguments based on observations \citep{mamajek}, and on theoretical work \citep{alexander2006}.

Photoevaporation of circumstellar disk gas may also be driven by far-ultraviolet (FUV) photons from the largest stars in the local star forming region. Classically, it is thought that FUV can only result in substantial mass loss if the outer disk boundary $r_d$ extends beyond the critical radius. We note that this radius is beyond the region of the solar system's ice giants. However, \citet{adams2004} found that FUV may be important at disk distances of $\sim 5-10$ times less than $r_g$. Conceptually, a sub-sonic gas outflow originates at $r_d$, which then flows through a sonic point (being accelerated to the sound speed), eventually reaching $r_g$ and escaping. Assuming external irradiation, \citet{adams2004} found that FUV may drive outflows at the current locations of the ice giants, up to stellocentrc distances of $\simeq 15$ AU. For example, the mass-loss rates due to photoevaporation by FUV photons are $ 10^{-9}$ to $10^{-8} M_{\Sun}/\text{yr}$ for a strongly FUV-irradiated disk at respective disk radii of 20 to 40 AU (see Fig. 6 of \citet{adams2004}). In this case, the gas disk depleted by a factor of 100 disperses on a timescale of: $\tau_\text{disp} \equiv M / \dot{M} \simeq 10^4 \text{ to } 10^5$ years.

X-rays may play an important role in determining the gas disk lifetime. X-rays heat the gas to $T \sim 10^3-10^4$ K, which corresponds to a critical radius for a solar-mass star of 10-100 AU. Owen et al. (2011) analytically estimated the mass-loss rate to be $\sim 8 \text{x} 10^{-9} \msun \text{yr}^{-1}$ for a Solar-mass star with an X-ray luminosity of $10^{30}$ erg $\text{s}^{-1}$ which would give a lifetime of $\sim 10^4$ yr for our depleted disk.

For completeness, we consider the timescale for viscous accretion of gas onto the Sun. Considering the model of viscous disk accretion in a disk with turbulence parameterized by the $\alpha$ parameter \citep{shakura}, the timescale for gas accretion in the nebula is:
\begin{equation}
\tau_\text{acc} = \frac{1}{\alpha \Omega}\left(\frac{H}{r}\right)^{-2}.
\end{equation}
We note that this timescale is independent of disk surface density. Using a passive disk temperature profile, and an $\alpha$ value\footnote{Values of $10^{-3}$ to $10^{-2}$ are often used partly because they give the approximate observed disk lifetimes} of $10^{-3}$, this timescale is $\sim$ 1-2 Myr on orbital separations of 20-40 AU.

 \section{Proposed Solution}
We suggest that the fine tuning problem may be resolved by proposing a dynamical reason for the ice giants to accrete their gas envelopes in the last $\lesssim 10^5$ years of the gas disk lifetime. Why did they start accreting their atmospheres precisely at that time?
In Section 3, we calculated that the ice giants have to accrete the bulk of their masses in gas from a disk depleted to $10^{-3}-10^{-2}$ times its original surface density, for their current atmospheric masses to be equal to the mass of gas in their respective feeding zones. Since the dust-to-gas ratio is $\sim 0.01$ in a full gas disk, in a depleted disk it's order unity: $\Sigma_{s}/\Sigma_{g} = \Sigma_{s}/\Sigma_{g,0} \text{ x } (\Sigma_{g}/\Sigma_{g,0})^{-1} \simeq 1 $.

A full gas disk efficiently damps the random velocities of protoplanetary cores, forcing them to stay on roughly circular orbits. However, once the gas is depleted to a point where the surface density of the solids becomes comparable to the surface density of the gas, gas drag can no longer balance dynamical excitations by neighboring bodies, causing the growing cores' random velocities to increase. The resulting dynamical instability may cause the cores to travel to a new location, continue their core growth there, and accrete a gas envelope from the remaining gas.

 \begin{figure}[h!]
 \centering
   \includegraphics[width=0.47\textwidth]{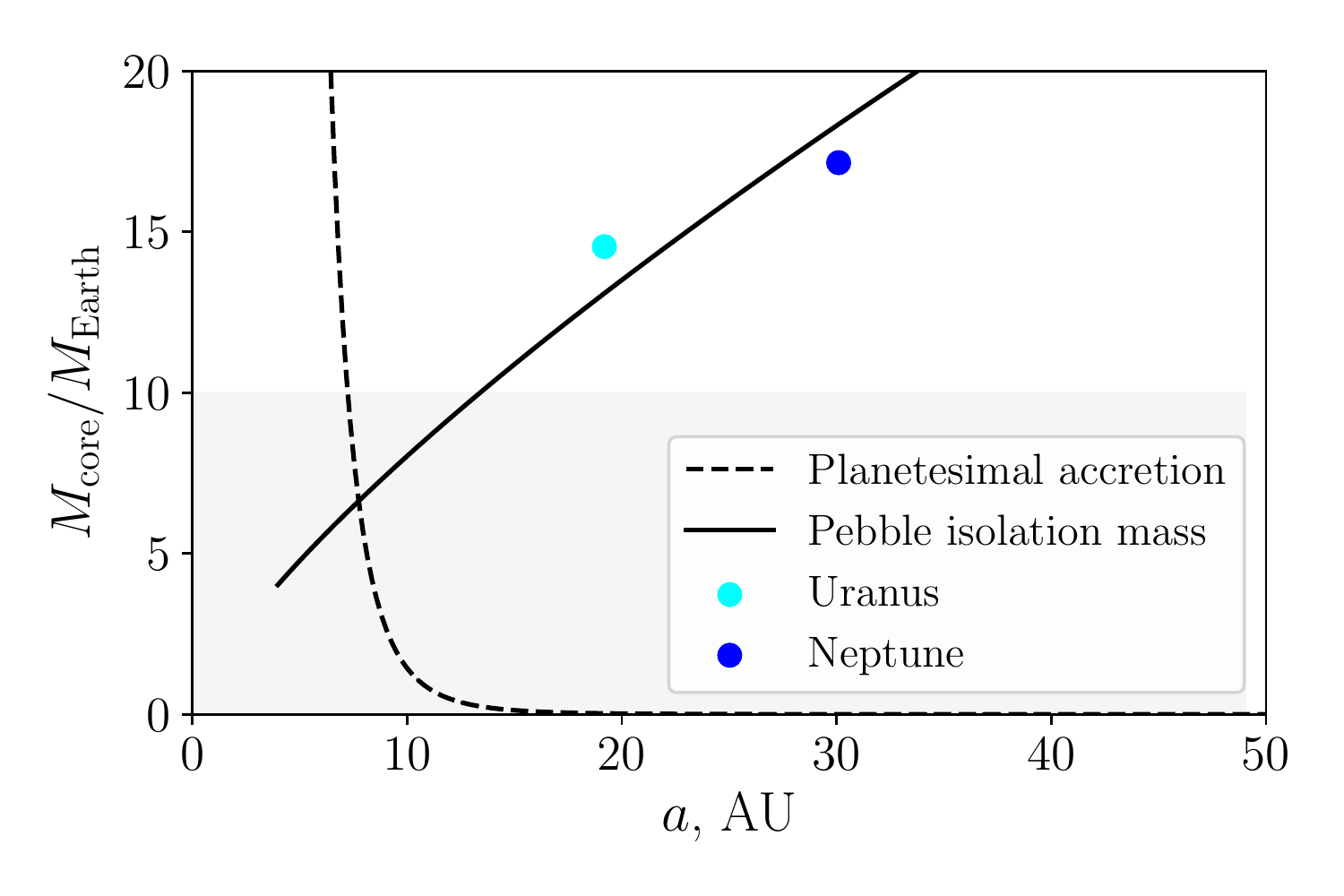}
   \caption{How massive can a core grow as a function of stellocentric distance (outside the terrestrial planet region)? Here, we speculate that pebble isolation sets the maximum core mass at distances comparable to those of the ice giants. Closer in, standard planetesimal accretion may set the core size. The maximum core mass that can be achieved by accreting planetesimals at random velocities of order the Hill velocity is shown (dashed line, Equation \ref{standard_growth}). This mass is limited by the supply of planetesimals, so plotted here is the (extreme) upper limit for standard growth. The solid line marks the pebble isolation mass (Equation \ref{pebiso}, scaled by a factor of 1/2), which is an upper limit for core growth via pebble accretion. The shaded region below 10 $M_\mathrm{Earth}$ represents (roughly) the scale at which cores are sub-critical.}
 \end{figure}
 
Why might the ice giants have remained small until this late stage? We suppose that the ice giants' cores started growing on closer-in orbits, where they were prevented from reaching a critical core mass. There are several possibilities that could have halted core growth on closer-in orbits. The differing turbulence structure in the disk could have affected pebble accretion. Alternatively, the ice giants could have formed in between Jupiter and Saturn, with Saturn consuming the infalling pebbles. Proximity to the gas giants, which excite the random velocities of neighboring bodies, can slow growth by accretion of large planetesimals. Finally, as the pebble isolation mass scales with stellocentric distance, the cores could have reached a sub-critical pebble isolation mass on closer-in initial orbits, if pebble accretion was the dominant mode of core growth at their initial orbital locations. We note that the core growth at smaller stellocentric distances (such as that of Jupiter and Saturn) is not limited to the pebble isolation mass, as significant accretion of km-sized planetesimals via the standard \citet{gls} regimes can occur within the gas disk lifetime. 

Plotted in Figure 3 are estimates on limits of core growth as a function of stellocentric distance, where the dominant accretion regime can vary depending on the orbital timescale. The mass a core can grow to within the disk lifetime $\tau_\text{disk}$ by accreting planetesimals at a rate $\dot{M_s}$ is: $M_\text{core,max} = \dot{M_s}\tau_\text{disk}$. Using the \citet{gls} growth rate for gravitational focusing at random planetesimal velocities of order the Hill velocity,
\begin{equation}
\dot{M_s} \simeq \Sigma_{s}(a) \Omega R_{p}^2 \left( \frac{v_\text{esc}}{v_H}\right)^2,
\end{equation}
where $\Sigma_{s}(a)$ is the disk surface density in solids at a stellocentric distance $a$, $\Omega=(G\msun/a^3)^{1/2}$ is the local Keplerian frequency, $R_p$ is the protoplanetary radius, and $v_\text{esc} = (2G\mcore/R_p)^{1/2}$ is the escape velocity at its surface, we solve for the maximum core size that a protoplanet of a mean density $\rho_c$ can achieve within the disk lifetime by accreting planetesimals:
\begin{equation} \label{standard_growth}
M_\text{core, max} \simeq \frac{54}{\pi} \frac{ G^{3/2} \msun^{1/2} \tau_\text{disk}^3 \Sigma_{s}(a)^3 }{ \rho_c a^{3/2} }.
\end{equation}
As the normalization of the pebble isolation mass is uncertain, we plot Equation \ref{pebiso} scaled by a factor of 1/2 for the masses to be comparable to the current ice giants' masses. To make this plot, we chose disk surface density and temperature profiles appropriate for the MMSN, a gas disk lifetime of 3 Myr, and a mean density $\rho_c = 3$ $\text{g}/\text{cm}^3$ for the core.

At distances close to the star, cores can grow efficiently via standard planetesimal accretion: there is no timescale problem as there is in the outer disk, and the core mass is limited instead by the supply of planetesimals. In the outer disk, the dynamical timescales are too long to form giant planets with standard growth mechanisms without appealing to accretion from a very thin planetesimal disk. However, as long as a core can grow to a fraction of an $\mearth$ required for pebble accretion to begin, it can grow to its pebble isolation mass very quickly. We observe in Figure 3 a portion of the plot where the maximum core masses falls below the nominal value for the critical core mass of 10 $\mearth$. This dip in the plot illustrates that there can be a region in the disk where gas giant cores cannot form in situ, surrounded by an area where they can.

We have discussed above several means by which the cores could have stayed sub-critical until onset of the dynamical instability in the depleted disk, bearing in mind that any model of solar system formation must keep Uranus and Neptune sub-critical during the full disk lifetime.

Once the gas was depleted enough to allow the cores' orbits to be gravitationally excited and not damped, the gas giants flung them outward, either to their current locations, or to a sufficient distance that they began migrating outward due to planetesimal-driven migration. Accreting the pebbles at their new locations allowed the cores to at least double in size to their final masses. Each core's orbit was circularized as a result of accreting a mass of roughly circular-orbit solids (the pebbles are coupled to the gas) comparable to its own mass.

Under the most favorable conditions for pebble accretion, from Equation \ref{pebble} the timescale for last doubling is $t_\text{growth} = M/\dot{M}$, or approximately $10^5$ years for a Neptune-sized core at its current location, with a surface density in solids given by the MMSN. We note that pebble accretion continues to operate in a depleted gas disk, though the sizes of pebbles that are effectively accreted change \citep{rosenthal2017}. The depleted disk dissipates on a $\lesssim 10^5$ year timescale due to EUV- and FUV- driven mass loss. Our depleted disk may dissipate on a shorter timescale of $10^4$ years due to photoevaporation driven by stellar X-ray radiation. If this shorter timescale applies, the cores of Uranus and Neputune may nonetheless undergo their last doubling in a timescale comparable to the remaining disk lifetime if they accrete from a disk enhanced in solids by a factor of only a few compared to the MMSN. Whether pebble accretion can decrease core growth timescales sufficiently to allow Neptune-sized cores to double within $10^4-10^5$ yr depends on the size distribution of planetesimals and the level of turbulence in the depleted gas disk (see \citet{rosenthal2017} for a detailed discussion). Here, we merely note that fast enough growth is possible. 

Such a high accretion rate limits the atmospheric mass that a core can sustain in a steady state \citep{rafikov2011}, possibly to a fraction of an Earth mass for the ice giants. For our scenario to work, the core accretion luminosity cannot be so high during the entire phase of last doubling that it prevents the accumulation of a couple $\mearth$ of gas.

Because the last doubling and disk dispersal timescales are comparable, it is reasonable to suggest that the ice giants' cores could have finished their growth as they accreted gas envelopes from the remaining gas disk. If the ice giants' cores doubled their mass while finishing their growth at their final orbital locations, their growth rate during this final doubling phase is approximately: $\dot{M} \simeq 6\mearth/(10^5 \text{yr}) \simeq 10^{16} \text{g}\text{ s}^{-1}$. \citet{rafikov2011} estimates the critical core mass in the fast planetesimal accretion regime as the mass of the core that can hold onto its own mass in gas. With masses of around $2 \mearth$, the atmospheres of Uranus and Neptune are not massive enough to undergo nucleated instability. We claim that the cores need not be super-critical to maintain $\sim 2 \mearth$ envelopes in a steady-state during rapid planetesimal accretion: a lower core mass is sufficient. Assuming that the planetesimal accretion luminosity is transported out via radiative diffusion through the atmosphere with a constant gas opacity 
$\kappa_0$, \citet{rafikov2011} finds that:
\begin{equation}
M_\text{atm}\simeq \zeta \left(\frac{GM_p\mu}{k}\right)^4\frac{\sigma}{\kappa_0L},
\end{equation} where $L$ is the accretion luminosity, $R_p$ is the radius of the core, and $\zeta = \zeta_0 M_p^{\delta}$, with $\zeta_0$ and $\delta$ as defined for Equation \ref{m_crit}. Therefore, a core accreting at a rate $\dot{M} \simeq 10^{16} \text{g}\text{ s}^{-1}$ at a distance $a=20\text{AU}$ can hold onto a $2 \mearth$ atmosphere as long as $\mcore \gtrsim 12 \mearth$. Thus, the cores of Uranus and Neptune in the final stage of growth were able to capture all of the gas in their feeding zones, forming steady state envelopes heated by fast planetesimal accretion. We note that the evolution timescale (given by the KH timescale) to such a steady state can be short, as a consequence of the high planetesimal accretion luminosity. If we suppose that the dominant source of the luminosity is the accretion luminosity ($L = GM_p \dot{M} /R$), then the KH timescale can be approximated as follows:
\begin{equation}
\tau_{KH} \simeq \frac{E}{L}\simeq \frac{GM_\text{p}M_\text{atm}}{R L} \simeq \frac{M_\text{atm}}{\dot{M}}.
\end{equation}
Therefore, for a $2 \mearth$ atmosphere surrounding a core accreting at a rate $\dot{M} \simeq 10^{16} \text{g}\text{ s}^{-1}$, the evolution timescale is approximately $\simeq 3 \text{x} 10^4$ yr. Additional effects such as the pollution of the envelope from the disruption of icy accreting solids have been shown to decrease this timescale substantially \citep{2015A&A...576A.114V}. If the appropriate timescale for the disappearance of the depleted disk is $10^4$ years, then the rate of pebble accretion (and, therefore, the luminosity) would have to be higher by a factor of 10 from the scenario considered above. In that case, according to Equation 19 the cores would only be able to hold onto $\sim 0.2$ Earth masses in gas, and we would require fast accretion to be halted before the dispersal of all of the disk gas.

We've proposed a dynamical scenario in which the ice giants would complete their core growth and accrete their gas envelopes in the time that the depleted disk dissipates. We emphasize that even if this scenario is incorrect, pebble accretion introduces a fine tuning problem in the standard scenario for the formation of Uranus and Neptune.

\section{Summary and Discussion}
Recent work on pebble accretion successfully resolves the timescale problem in growing gas giant cores at large orbital distances, but it gives rise to a fine tuning problem in the formation of the ice giants. In the standard scenario, the majority of solid accretion halts once the cores reach a certain mass, at which point the cores slowly begin accreting significant amounts of gas. However, pebbles, which are mm- to cm- sized particles observed to exist in the outer regions of protoplanetary disks, accrete extremely efficiently onto protoplanetary cores at the mass scale of the ice giants. They accrete so quickly that it is difficult for ice giant cores to exist at roughly their current core masses for a significant fraction of the disk lifetime, while accreting a gas envelope. To avoid runaway, they must therefore complete their growth at a specific time, when the gas disk is partially, but not entirely, depleted. Halting core growth at a specific point in time is fine tuned, and in this paper we propose a scenario for resolving this problem of fine tuning in the growth of Uranus and Neptune.




We suggest that the onset of the majority of gas accretion onto the planetary cores coincides with a dynamical instability, where the gas was no longer able to damp the growing cores' random velocities. We suppose that the cores of Uranus and Neptune could have begun their growth closer to the gas giants. Then, at the point where $\Sigma_\text{solid} \sim \Sigma_\text{gas}$, their orbits were gravitationally excited and they may have traveled outward to their present locations. There, the ice giants' orbits circularized, and they accreted their $\sim 15$\% atmospheres from the now-depleted nebula, avoiding atmospheric runaway.

We emphasize that the fine tuning problem exists regardless of the explanation. Since they never underwent runaway, the ice giant cores could not have grown to their current masses in a full gas disk, in which they can grow rapidly by accreting pebbles. As an alternative to our scenario, the cores may halt their growth if:

\begin{itemize}
\item there is no radial drift of pebbles into the ice giants' respective feeding zones. This is improbable, given protoplanetary disk observations of pebbles;

\item the supply of pebbles in the outer disk is somehow cut off, which would require a cutoff mechanism;

\item core masses are limited by pebble isolation \textit{and} the grain opacity is higher than expected (such that $M_\text{iso,peb}<\mcrit$ for the ice giants). This high opacity is unlikely, given the existence of the solar system's gas giants.
\end{itemize}

We predict that the occurrence of isolation-mass scale ice giants beyond 10 AU with ~0.1 atmospheres by mass may be strongly correlated with the presence of gas giants in the system. If our proposed mechanism is correct, then gas giants---by facilitating substantial migration during the epoch of instability---may play a key role in forming planets with intermediate-size atmospheres in the outer disk.  Without gas giants, planets likely won't move around as much at our preferred time, so they are less likely to experience enhanced growth while the gas disk is dissipating. They could still grow to the scale of the ice giants after the gas disk is gone, but in that case, they would have small rather than intermediate-sized atmospheres. Therefore, we predict that when gas giants are not present, outer planets will host smaller atmospheres.

We are thankful to Jonathan Fortney, Jack Lissauer, Michael Rosenthal, and Yanqin Wu for helpful discussions. We thank the referee for comments that improved this paper. This work was supported by NSF Career grant number AST-1555385.

\bibliographystyle{aasjournal}

\end{document}